\newcount\style \style=1
\ifodd\style
  \documentstyle[aas2pp4]{article}
\else
  \documentstyle[12pt,aasms4]{article}
\fi %\input psfig

\def\del{\nabla}
\def\bcdot{\cdot}
\def\btimes{\times}
\def\dd{\partial}

\def\HI{\ion{H}1}
\def\etal{et al.}
\def\eg{e.g.}
\def\etc{{\it etc.}}
\def\ie{i.e.}
\def\beq{ \begin{equation} }
\def\eeq{ \end{equation} }
\def\spose#1{\hbox to 0pt{#1\hss}} % from Scott Tremaine
\def\ltsim{\mathrel{\spose{\lower.5ex\hbox{$\mathchar"218$}}
     \raise.4ex\hbox{$\mathchar"13C$}}}
\long\def\Ignore#1{\relax}

\begin{document}

\title{Differential Rotation and Turbulence in Extended \HI\ Disks}

\author{J.\ A.\ Sellwood\footnotemark[1]}
\affil{Department of Physics and Astronomy, Rutgers, The State University
of New Jersey, 136 Frelinghuysen Road, Piscataway, NJ 08854-8019}
\affil{sellwood@physics.rutgers.edu}

\author{Steven A.\ Balbus\footnotemark[1]}
\affil{Virginia Institute of Theoretical Astronomy, Department of
Astronomy, University of Virginia, Charlottesville, VA 22903-0818}
\affil{sb@virginia.edu}

\addtocounter{footnote}{1}
\footnotetext{\it Also, Isaac Newton Institute for Mathematical Sciences,
Cambridge University, 20 Clarkson Rd., Cambridge CB3 0EH, UK}

\begin{abstract} 

When present, extended disks of neutral hydrogen around spiral galaxies
show a remarkably uniform velocity dispersion of $\sim 6$ km s$^{-1}$.
Since stellar winds and supernovae are largely absent in such regions,
neither the magnitude nor the constancy of this number can be accounted
for in the classical picture in which interstellar turbulence is driven
by stellar energy sources.  Here we suggest that magnetic fields with
strengths of a few microgauss in these extended disks allow energy to be
extracted from galactic differential rotation through MHD driven turbulence.
The magnitude and constancy of the observed velocity dispersion may be
understood if its value is Alfv\'enic.  Moreover, by providing a simple
explanation for a lower bound to the gaseous velocity fluctuations, MHD
processes may account for the sharp outer edge to star formation in galaxy
disks.

\keywords{galaxies: ISM --- galaxies: kinematics and dynamics ---
hydrodynamics --- instabilities --- radio lines: galaxies --- turbulence}

\end{abstract}

\section{Introduction}

It has long been recognized that the interstellar medium (ISM) in galactic
disks is turbulent (\eg\ Scalo 1987, Dickey \& Lockman 1990).  Within our
own Galaxy, line widths of individual molecular and \HI\ clouds, and
cloud complexes greatly exceed the expected thermal width, when kinetic
temperatures can be estimated (Miesch \& Bally 1994; Caselli \& Myers 1995).
There is also evidence for a high velocity dispersion population of \HI\
clouds in both the Milky Way Galaxy (Radhakrishnan \& Srinivasan 1980,
Kulkarni \& Fich 1985), and for a high velocity tail in the gas where
the disk is optically bright in some external galaxies (Dickey, Hanson,
\& Helou 1990; Boulanger \& Viallefond 1992; Kamphuis \& Sancisi 1993;
Schulman \etal\ 1996; Braun 1997).

The classical picture (\eg\ Spitzer 1978) is that supernovae and other
stellar processes (\eg\ winds, outflows, \etc)\ supply the requisite energy
to maintain turbulent cloud velocities against dissipative losses.  In the
absence of such sources, one would therefore expect interstellar turbulence
to decay rapidly.  In particular, turbulent motion should be negligible
beyond the outer optical edges of spiral galaxies, where there are few stars.

This expectation is not in accord with observations.  Radio observers
report that 21 cm dispersions do not drop below 5 -- 7~km~s$^{-1}$,
regardless of how low the optical surface brightness becomes (\eg\ Dickey
\etal\ 1990; Kamphuis 1993).  If this line width is a measure of
the turbulent motion in an ensemble of cool clouds, the medium is being
stirred by other means.  We here propose that MHD instabilities (Balbus \&
Hawley 1991) in the differentially rotating gas layer are responsible for
a minimum level of turbulence in all gas disks.

The need for MHD mediation of interstellar turbulence has been noted by a
number of other investigators.  Magnetic fields in galactic disks tend to
be partly tangled on small scales, where lines of force have been observed
to thread and be compressed by dense clouds, but also to retain some
large-scale coherence (Heiles \etal\ 1993; Heiles 1996).  On the largest
Galactic scales, the field energy density is less than that of the random
cloud motions, but probably in excess of the average thermal energy density.
Within molecular cloud complexes, the field strength can be dominant,
and MHD processes in these systems are central to understanding their
dynamics and evolution (\eg\ McKee \& Zweibel 1995; Gammie \& Ostriker 1996).

The extended \HI\ disks of some galaxies are a relatively clean laboratory
in which interstellar turbulence can be observed on its largest scales.
It is clearly present in these systems, even though star formation and
the associated supernova rates are very low.  If the turbulent motions
in this regime are Alfv\'enic, we show that they are maintained through
magnetically mediated dynamical heating from differential rotation.
We present a derivation of this heating rate when magnetic fields {\it and\/}
self-gravity are both present, though in our specific application the process
is magnetically dominated.  Our result is applied to the particular case
of NGC 1058, a face-on disk galaxy with a well-studied extended \HI\ disk.
Both the constancy and magnitude of the observed velocity dispersion follow
naturally from an MHD explanation.

\section{Turbulence in extended \HI\ disks}

A compilation of high-quality neutral hydrogen line width data on several
galaxies may be found in Kamphuis (1993).  The principal conclusions from
this study are that 21 cm dispersions are $\sigma \sim 10 - 12$ km s$^{-1}$
in the bright optical disk (with evidence for still higher values in spiral
arms), and that the dispersions decline to no less than $5-7$ km s$^{-1}$
when the optical surface brightness falls below 25 B mag arcsec$^{-2}$.
The higher spatial resolution data presented by Braun (1997) indicate
that broad lines are found in a ``high brightness network'' absent in the
optically faint outer disks.

Dickey \etal\ (1990) drew particular attention to the remarkable lack
of spatial variation of the \HI\ line width in the gas layer beyond the
optical disk in NGC 1058.  Data from other galaxies show a similar trend,
and indicate that the velocity dispersion puzzle is widespread.  Both NGC
1058 and NGC 5474 (Rownd \etal\ 1994) are almost face-on, allowing clean
measurements of line widths with little contribution from beam smearing.
The velocity dispersions of these galaxies appear to approach a steady value
of $\sigma \sim 6 \hbox{ km s}^{-1}$ at radii outside the bright disk.
(The missing flux in their observations could, in principle, affect this
result, but a similar dispersion is obtained for other galaxies (\eg\
Kamphuis 1993) in which the full flux is recovered.)  Dickey \etal\
stress that where the line width is small, the \HI\ line profiles are
almost perfect Gaussians.

Let us focus now upon NGC 1058, a small galaxy with an adopted distance
of 10~Mpc.  Its optical radius $R_{25} = 90^{\prime\prime}$ or 4.5 kpc
and the \HI\ layer extends to $220^{\prime\prime}$ or 11 kpc. Since it
is almost face-on, its rotation curve is poorly known; we adopt a flat
circular velocity of 150 km s$^{-1}$, consistent with the \HI\ velocity field
with $0.05 \ltsim \sin i \ltsim 0.1$.  Dickey \etal\ report a \HI\ column
density $\sim 3 \times 10^{20} \hbox{ cm}^{-2}$ at $R\sim5$~kpc falling to
$2 \times 10^{20} \hbox{ cm}^{-2}$ at $R\sim 10$~kpc.  These are likely to
be underestimates, however, since their interferometric data miss about half
the flux determined from single dish observations.  Therefore, as fiducial
values, we double the column density numbers in our calculations below,
adjust for helium content and also assume there is no significant quantity
of molecular gas in the outer layer.  We adopt a total thickness of the
outer \HI\ layer of $h = 400$ pc -- about twice the thickness of the gaseous
disk of the Milky Way in the Solar neighborhood, since we expect the layer
to flare to some extent.  We summarise the properties we adopt in Table 1.

The constant internal \HI\ line width, Dickey \etal\ argue, could represent
either the thermal temperature of the gas or the turbulent motion of many
small, cooler clouds.  Neither interpretation is free of puzzles.

If the observed 1-D dispersion of $\sim 5-7$ km s$^{-1}$ is thermal, it
would imply a gas temperature of $\sim 3\,000-6\,000$~K, perhaps less if
there are some contributions to this width from beam smearing (small in a
nearly face-on galaxy) and turbulence.  Dickey (1996) points out that atomic
hydrogen at a temperature of a few thousand degrees would be in an unstable
state.  Braun (1998) expects the low ambient pressure to cause all gas to
be in the warm phase at a temperature of $\sim 10\,000$~K ($\sigma \sim 9
\hbox{ km s}^{-1}$), which is inconsistent with the observed dispersion.
While the principal heating and cooling processes at work in the Solar
neighborhood (Wolfire \etal\ 1995) are likely to be different in the far
outer disk, they would have to be drastically so for gas to be in a stable
phase at $\sim 5\,000$~K.  Moreover, stars {\it are\/} forming at a very
low rate in at least some of these gas layers (Ferguson \etal\ 1998), which
must imply the existence of cooler, denser gas from which they can form.

We therefore prefer to attribute the observed line width largely to
turbulent motion in an ensemble of small, cool clouds, which immediately
raises two related questions: i) What process accelerates the clouds,
maintaining velocities well in excess of the internal sound speed of
individual clouds?; and ii) Why should the turbulent velocities be so
uniform?  We offer answers to both these questions in \S4.

\subsection{Supernova heating}

Supernovae (SNe) are generally assumed to be the dominant energy source for
the interstellar turbulent cascade (\eg\ Spitzer 1978; Norman \& Ferrara
1996).  The stellar density and supernova rate (SNR) in the gas layer
beyond the Holmberg radius is clearly much lower than in the bright inner
parts of galaxies, but could it still be sufficient to drive the turbulence?

Deep CCD images of NGC 1058 (Ferguson \etal\ 1998) have confirmed the
very faint outer spiral arms, first detected by Tamman (1974) in his
search for the stellar population responsible for the two SNe detected
during the 1960s in the outer parts of this galaxy.  Ferguson (1997)
estimates an azimuthally averaged surface brightness declining to 28 B
mag arcsec$^{-2}$, with a regular spiral pattern.  Ferguson \etal\ also
detect small \ion{H}2 regions along these spiral arms, but the inter-arm
region seems to be almost devoid of any star formation.

From the H$\alpha$ flux, Ferguson (private communication) estimates the
azimuthally averaged star formation rate (SFR) per unit area at $\sim$ 1.5
times the optical radius to be $\sim 5\times10^{-11}\hbox{ M}_\odot\hbox{
pc}^{-2}\hbox{ yr}^{-1}$.  A high estimate for the SNR for this SFR is
$10^{-12}$ SN pc$^{-2}$ yr$^{-1}$ (\eg\ Leitherer \& Heckman 1995).
Assuming mechanical energy input of $\epsilon 10^{51}$ ergs per SN,
where $\epsilon$ is the highly uncertain efficiency factor, we obtain
an energy input rate of $\sim \epsilon 3\times10^{-27} (400{\rm pc}/h)$
ergs cm$^{-3}$ s$^{-1}$.

For $\epsilon \sim 0.01$ (Chevalier 1998, private communication), this
energy input rate, while highly uncertain, is not decisively less than
that of the differential rotation source we consider below.  Neverthless,
it seems unlikely to be the principal source of cloud motions, because
there is no correlation whatsoever between the SFR and \HI\ velocity
dispersions across the extended disk.  Dickey \etal\ (1990) stress the
uniformity of the dispersion, while Ferguson \etal\ observe \ion{H}2\
regions to lie almost exclusively in very narrow arms.

Braun's (1997) high spatial resolution studies of \HI\ in the bright
inner parts of other galaxies show a ``high brightness network'' of
distinctly non-Gaussian line profiles with broad, high velocity tails.
He further reports (Braun 1998) a loose correspondence between the network
of extended \HI\ linewidths in M31 and the observed H$\alpha$ emission,
and argues for a causal connection between linewidth and energy injection
from stellar activity.  Thus turbulence driven by stellar activity appears
to create broad wings in the line profiles in localized regions where the
energy is deposited.  It is hard to see how this activity could also produce
the observed uniform level of near Gaussian line profiles on larger scales.

\subsection{Other sources of turbulence}

Other sources of turbulence can be imagined.  Infall models include the
returns from a galactic fountain (Bregman 1980; Schulman \etal\ 1996), or
chimney (Norman \& Ikeuchi 1989), or simply the direct accretion of external
intergalactic matter (T\'oth \& Ostriker 1992; Kamphuis \& Sancisi 1993).
Gravitational scattering by transient spiral waves (Carlberg \& Sellwood
1985; Jenkins \& Binney 1990; Toomre \& Kalnajs 1991) is yet another
possible energy source.

Dickey \etal\ (1990) already remarked that the very low SFR in NGC~1058
is almost certainly inadequate to drive a vigorous galactic fountain.
Even granting the presence of a weak fountain, the measured \HI\ dispersion
is no lower than in the outer parts of other galaxies with considerably
more active star formation (Kamphuis 1993).  Furthermore, stirring of the
\HI\ layer by infalling dwarf galaxies and debris is unlikely to maintain a
uniform level of mild turbulence everywhere, and should result in radiative
emission of a substantial fraction of the infall energy.

Transient spiral waves produce a choppy potential sea which scatters
stars, and any other material in the disk, causing random motion to rise.
Where most of the disk mass is in the collisionless component -- \ie\
the stars -- the process is self-limiting, since the spiral waves quickly
become weaker as the velocity dispersion of the stars rises.  Because gas is
able to dissipate turbulent energy through inelastic collisions, its fate
will be different, however.  It seems unlikely that gas would settle to a
smooth distribution on large scales with a level of turbulence resulting
from a balance between scattering and dissipation; it is more likely that
stellar spiral arms promote the formation of large gas concentrations
in which further gravitational instability will lead to star formation.
In parts of the disk in which gas is the dominant mass component, if
gravitational instability is present on large scales, we might expect it
to cascade directly to forming stars.

The importance of self-gravity in a disk may be determined by evaluating
the usual local stability parameter (\eg\ Binney \& Tremaine 1987)
 \beq
Q = {\sigma \kappa \over \pi G \Sigma},
 \eeq
where thickness corrections have been neglected.  Here $\sigma$ is the
velocity dispersion, the epicyclic frequency $\kappa = \sqrt{2} V_{\rm
circ}/R$ for a flat rotation curve, and $\Sigma$ is the surface density.
Values of this parameter for the axially symmetrized gas only in the outer
\HI\ layer of NGC 1058 are evaluated in Table 1.  The $Q$ value for the
extremely faint stellar disk seems to be huge: $Q \sim 60 (6\hbox{ kpc}/R)
(\sigma_u/10\hbox{ km s}^{-1}) / \Upsilon_B$, where $\Upsilon_B$ is the
mass to B-band luminosity in Solar units; any supporting response from
the stars must be utterly negligible in this region.

The values of $Q$ given in Table 1 are uncertain and rather modest to argue
strongly that the gas layer is clearly stable.  We note, however, that
the gas surface density varies by perhaps a factor two between the peaks
in the spiral arms and the inter-arm level.  Ferguson \etal\ (1998) report
mild star formation, a clear indicator of local Jeans instability, in the
arms but the paucity of detectable star formation between the arms suggests
that the layer is stable there.  The origin of the spiral arms in this outer
layer is unclear, but gravitational instability seems quite unlikely since
the Jeans length ($\lambda_{\rm crit} = 4\pi^2G\Sigma_{\rm gas}/\kappa^2$)
is much too small for such a large-scale and symmetric two-armed spiral
(Table 1).  We conclude that gravitational instabilities are unlikely to
be stirring the gas layer to maintain the low level of turbulence in the gas.

\section{Turbulent heating by coupling to differential rotation}

In this section we consider the local turbulent dynamics of a galactic
disk.  Disk material orbits in the global potential of the galaxy, but is
locally subject both to a magnetic field and (in principle) to its own
self-gravity.  We require the form of the turbulent volumetric heating
rate under these conditions.  For the reasons just given, self-gravity
is not of direct importance for the gaseous disks under consideration
here but we include it in the analysis of this section for the sake of
completeness, and because it is an interesting problem in its own right.
In the presence of self-gravity, the volumetric turbulent heating rate
of the gas due to differential rotation is a direct generalization of the
nonself-gravitating case: a coupling between the effective stress tensor
and the large scale angular velocity gradient.

In a homogeneous gas, if the field is subthermal, the free energy of
differential rotation drives a dynamical instability (Balbus \& Hawley 1991,
1998), extracting rotational energy and depositing it in turbulent motions.
The physical mechanism for the generation of turbulent motion is clear.
Mass elements orbiting in the fluid at slightly differing radii, but coupled
magnetically, pull on each other as the shear attempts to separate them as if
connected by a weak spring.  The effect of their mutual forces is to remove
angular momentum from the one which has less and donate it to the one which
has more.  As a result, the radial separation of the elements increases,
which increases the difference in angular velocity and the instability
runs away -- provided the spring is not strong enough to resist.  We have a
rather different case in mind here: a highly inhomogeneous gas in which the
relevant ``thermal'' motions refer to the macroscopic velocity dispersion.
It is likely that random magnetic stresses will continue to tap into the
differential rotation as a source of turbulence for relatively strong
fields as well (e.g.\ Eardley \& Lightman 1975).  Indeed, the expected
outcome of differential rotation and any radial magnetic field component
is a positive radial-azimuthal Maxwell stress; this alone is sufficient
to drive noncircular velocity fluctuations (cf.\ below).

Adopting a standard $(R, \phi, z)$ cylindrical coordinate system, we denote
the circular velocity as $R\Omega\hat{\bf e}_\phi$, where $\hat{\bf e}_\phi$
is a unit vector in the azimuthal direction.  The velocity vector ${\bf
u}$ is the difference between the true velocity and the azimuthal circular
velocity; it is a fluctuation velocity satisfying $u \ll R\Omega$.  We allow
for the possibility of a mean, slowly varying drift velocity, denoted
$\langle {\bf u} \rangle$, which is much less than the RMS fluctuation
$\langle u^2 \rangle^{1/2}$.  The angle brackets $\langle\ \rangle$ denote
local averages in radius and height, but a complete average in azimuth.  Thus
 \beq
| \langle {\bf u} \rangle | \ll \langle u^2\rangle^{1/2} \ll R\Omega.
 \eeq
The Alfv\'en velocity associated with the magnetic field ${\bf B}$ is
given by
 \beq
{\bf u_A} ={ {\bf B}\over \sqrt{4\pi\rho}}
 \eeq
where $\rho$ is the mass density.  We do not make a formal distinction
between the mean magnetic field and its fluctuating component, but like
its kinetic counterpart, the magnitude $u_A$ is assumed to satisfy $u_A
\ll R\Omega$.

We work in the standard {\it local approximation\/} in which $R$ is assumed
large enough that we can ignore geometrical curvature terms.  Our starting
point is the energy equation for the $u$ fluctuations with the gravitational
contribution written explicitly as a power term (see equation [89] in the
review of Balbus \& Hawley 1998):
 \begin{eqnarray}\label{energy}
\lefteqn { {\dd\ \over\dd t} \left\langle {\textstyle{1\over2}}\rho
(u^2+u_A^2 ) \right\rangle + \del\bcdot \langle \>\rangle = \qquad\qquad}
\nonumber\\
 & & - {d\Omega\over d\ln R} \left\langle\rho
(u_R u_\phi - u_{AR} u_{A\phi})\right\rangle \nonumber\\
 & & - \left\langle
 \rho {\bf u}\bcdot\del\Phi\right\rangle +
 \left\langle P\del\bcdot{\bf u}\right\rangle\nonumber\\
 & & - \sum_i \left\langle \rho \nu |\del u_i|^2 - {\eta\over 4\pi} |\del B_i 
|^2
\right\rangle
 \end{eqnarray}
Here $P$ is the gas pressure, $\nu$ the kinematic viscosity, $\eta$ the
resistivity and the operand in the divergence term $\del\bcdot\langle\>
\rangle$ is the energy flux
 \beq
\left( {\textstyle{1\over2}}\rho u^2 + P\right){\bf u} + { {\bf B} \over
4\pi} \btimes ({\bf u}\btimes {\bf B} ).
 \eeq
The gravitational potential $\Phi$ includes both a contribution from
self-gravity, plus a contribution from the external large scale potential.
Locally, the latter may be taken to depend upon $z$ only,
 \beq
\Phi=\Phi_{\rm sg} (R, \phi, z)\, {\rm (self\> gravity)} + \Phi(z)_{\rm ex}
\, {\rm (external)}.  \eeq

Equation (\ref{energy}) balances changes in the energy density against the
net flux, and explicit sources and sinks.  The first term on the right
side is the energy released by the Reynolds-Maxwell stress, and is the
key term in this paper.  The other terms are the power extracted from the
gravitational potential, the work done by pressure forces, and the viscous
and resistive losses.  Internal sources for interstellar energy fluctuations
(\eg\ SNe) can be included in the $P\del\bcdot{\bf u}$ source term.

We now show that the gravitational term can be manipulated into the form
of a coupling to the stress tensor.  Start with
 \beq\label{gravity}
\rho {\bf u} \bcdot\del \Phi = \del\bcdot(\rho{\bf u} \Phi)-
\Phi\del\bcdot(\rho{\bf u}).
 \eeq
Mass conservation and the Poisson equation then imply,
 \beq
-\Phi\del\bcdot(\rho{\bf u}) = \Phi \left( {\dd\ \over\dd t} + \Omega{\dd\
\over\dd\phi}\right){\nabla^2\Phi\over4\pi G}.
 \eeq
(N.B.\ The second term in the brackets arises because the total velocity
is ${\bf u} + R\Omega \hat{\bf e}_\phi$.)  Interchanging the order of
partial derivatives and integrating by parts leads to
 \beq
\Phi {\dd\ \over\dd t} {\nabla^2\Phi\over4\pi G} = {1\over 4 \pi G}
\left[ \del\bcdot\left(\Phi{\dd\del \Phi\over\dd t}\right) -{1\over2}
{\dd\ \over \dd t} \left|\del\Phi\right|^2\right].
 \eeq
Similar, but more lengthy, manipulations lead to
 \begin{eqnarray}
\lefteqn{ \Phi\Omega{\dd\ \over\dd\phi} {\nabla^2\Phi\over4\pi G} = {1\over
4 \pi G} \bigg[  \del\bcdot\left(\Phi\Omega{\dd\del\Phi \over\dd\phi}\right)}
\nonumber\\
 & & + \left(\dd\Phi\over\dd R\right) \left({\dd\Phi\over R\dd\phi}\right)
\left( d\Omega\over d\ln R\right)\bigg]\nonumber \\
 & & - {\dd \over\dd\phi}(\cdots)
\end{eqnarray}
The final partial $\phi$ derivative will vanish upon averaging, and is
not explicitly written.  Carrying this average through and returning to
equation (\ref{gravity}) leads to
 \begin{eqnarray}
\lefteqn{ \langle \rho {\bf u} \bcdot\del \Phi\rangle = -{\dd\ \over\dd t}
\bigg\langle{|\del\Phi|^2\over 8 \pi G} \bigg\rangle} \nonumber \\
 & & + \del\bcdot \bigg\langle \rho{\bf u}\Phi+ \Phi \left( {\dd\ \over\dd
 t} + \Omega{\dd\ \over\dd\phi} \right) {\del\Phi\over 4 \pi G} \bigg\rangle
 \nonumber \\
 & & + \langle\rho u_{GR} \, u_{G\phi}\rangle {d\Omega\over d\ln R}
 \end{eqnarray}
where the gravitational velocity ${\bf u_G}$ is defined by
 \beq
{\bf u_G} = {\del\Phi\over \sqrt{4\pi G\rho}}
 \eeq
If we now place our findings back into the energy equation (\ref {energy}),
we may write the result as
 \begin{eqnarray}
\lefteqn{ {\dd\ \over\dd t} \left\langle \textstyle{1\over2}\rho
(u^2+u_A^2-u_G^2 ) \right\rangle + \del\bcdot \langle \>\rangle = -
{d\Omega\over d\ln R} T_{R\phi}} \nonumber \\
 & & + \left\langle P\del\bcdot{\bf u}\right\rangle
- \sum_i \left\langle \rho \nu |\del u_i|^2 - {\eta\over 4\pi} |\del B_i
|^2 \right\rangle
 \end{eqnarray}
where the stress tensor $T_{R\phi}$ is defined as
 \beq \label{T}
T_{R\phi} \equiv \langle \rho( u_R u_\phi - u_{AR} u_{A\phi} + u_{GR}
u_{G\phi} ) \rangle
 \eeq
and the supressed energy flux has become
 \begin{eqnarray}
\lefteqn{ \left( {1\over2}\rho u^2 + P\right){\bf u} + { {\bf
B}\over4\pi}\btimes ({\bf u}\btimes {\bf B} ) + \rho{\bf u} \Phi} \qquad
\qquad  \ \ \ \ \nonumber \\
& & + \Phi \left( {\dd\ \over\dd t} + \Omega{\dd\ \over\dd\phi}
\right){\del\Phi\over 4 \pi G}.
 \end{eqnarray}
The utility of writing the self-gravity in the form of a stress tensor was
first noted by Lynden-Bell \& Kalnajs (1972). We see that the most general
stress tensor is a simple sum of the Reynolds, Maxwell, and Newtonian
stresses.  Free energy can be extracted from differential rotation at the
rate per unit volume $T_{R\phi}d\Omega/d\ln \,R$.

It should be noted that the instabilities, both MHD and gravitational,
drive the in-plane components of the velocity dispersion and do not couple
directly to motions normal to the plane.  Some degree of velocity anisotropy
is therefore expected in turbulence driven by these stresses.

\section {Application to NGC 1058}

Differential rotation supplies energy to fluctuations at a rate given by
 \beq
-T_{R\phi} {d\Omega\over d\ln R} = \Omega T_{R\phi}
 \eeq
for a flat rotation curve.  We do not know $T_{R\phi}$ {\it a priori,}
but numerical simulations of magnetized nonself-gravitating disks, carried
out under a wide variety of field geometries, equations of state, and
numerical grids consistently yield
 \beq\label{Tnum}
T_{R\phi} \simeq 0.6 \ {B^2\over 8 \pi}
 \eeq
(Hawley, Gammie, \& Balbus 1995), where the magnetic field is evaluated at
the time of saturation.  A somewhat delicate point is how to relate $B^2$
to the mean magnetic field.  The simulations reveal a field dominated by
its largest scales, and we shall assume that mean field and fluctuations
are comparable.  Assuming the outer disk in NGC 1058 is not self-gravitating
(as argued in \S2.2), equation (\ref{Tnum}) gives a simple, physically sensible
estimate for the stress tensor, which we shall adopt.

The scale for the fluctuating kinetic velocities is set by the Alfv\'en
velocity:
 \beq
\langle \rho u_R u_\phi \rangle \simeq  0.6 {B^2\over 8 \pi} + \langle
\rho u_{AR} u_{A\phi} \rangle
 \eeq 
Setting the left hand side to a fraction $f_1$ of $\overline{\rho} \sigma^2$,
where $\overline{\rho}$ is the mean density and $\sigma^2$ the measured
one-dimensional velocity disperion, and the entire right hand side to a
fraction $f_2$ of ${\overline {B}}^2/8\pi$, where $\overline{B}$ is the
mean magnetic field, we find for a velocity dispersion of 6 km s$^{-1}$
 \beq
\overline{B} = 3\> \mu{\rm G}\ \left(f_1\over f_2\right )^{1/2} \left(
{\overline \rho} \over 10^{-24} \hbox{ g cm}^{-3} \right)^{1/2}.
 \eeq
The implied field strength is a few microgauss -- a reasonable number for
a spiral galaxy (\eg\ Beck \etal\ 1996).  Observational confirmation of
this value would be challenging.
[The radio continuum emission from NGC 1058 is very weak (van der Kruit \&
Shostak 1984) because of its very low SFR.]

If we turn the problem around and ask how $\sigma$ should vary with
$\overline B$ and $\overline \rho$, as the density varies between and within
the spiral arms say, a natural explanation for its near constancy emerges.
Large scale but otherwise random field lines tend to follow a $\overline
B \propto {\overline\rho}^n$ scaling, by making the usual assumption of
flux freezing.  The index $n=2/3$ for isotropic spherical compression,
while $n=1$ for compression in a plane.  Since $\sigma$ is of order the
Alfv\'en speed, this implies
 \beq
\sigma \sim \left( {\overline\rho}\right) ^{n - 1/2}
 \eeq
\ie\ a rather weak dependence on density.  Therefore, a magnetic basis for
the velocity dispersions gives, in addition to a sensible inferred field
strength, a very simple basis for understanding the near constancy of
$\sigma$ in gas-dominated galactic disks.

Finally, let us estimate the energy deposition rate, using the numerical
result (\ref{Tnum}), and compare it with the SNR from \S 2.1.  Inserting the
above derived field strength of 3 $\mu$G, we obtain
 \beq
T_{R\phi} \Omega \simeq 1.4\times 10^{-28}\ {\rm ergs\ cm^{-3}\ s^{-1}}
 \eeq
at $R=7$ kpc.  Thus differential heating is certainly competitive with
the 1\% efficient SNR of \S 2.1, and provides a more natural explanation
for the observed uniformity of the turbulence.

\section{Summary}

We have argued that energy to drive turbulence in the ISM of a galaxy can be
extracted from the differential rotation either by way of a well-established
MHD instability (Balbus \& Hawley 1991), or directly from the $T_{R\phi}$
Maxwell stresses if the field is too strong to be formally unstable.
In the bright inner disk, where stars are forming and the SNR is high,
the classical picture of turbulence driven by stellar processes prevails,
but where the stellar density is low, the energy deposition rate from MHD
driven turbulence becomes important.

A quite reasonable field strength (a few microgauss) is required for the
turbulence to be Alfv\'enic, and its observed striking uniformity is,
we argue, a simple consequence of flux freezing.  But turbulence will
decay without a constant energy source.  We argue that MHD instabilities
are the preferred energy source in the outer \HI\ layer because they
are inevitable and will maintain the characteristic turbulent velocity
near the Alfv\'en speed.  Other sources of energy, SNe, infall, etc.\
are also available, and may be important locally, but do not provide a
natural explanation for the magnitude and uniformity of the turbulence.

By providing a mechanism for maintaining turbulence in the absence
of mechanical energy input from stellar processes, we can claim some
theoretical understanding of the semi-empirical star formation threshold
advanced by Kennicutt (1989).  He argues that (significant) star formation
in a galactic disk is truncated abruptly at the outer edge by the rising
value of $Q$, {\it assuming\/} the observed constant velocity dispersion
of 6~km~s$^{-1}$.  He shows that this simple criterion accounts for the
extent of most star formation in many galaxies.  Here, we are able to
provide a possible explanation for the constant velocity dispersion by
linking it directly with an Alfv\'en speed (cf.~eq.~[20]).  We are unaware
of any previous explanation for the maintenance of the observed turbulence
required for Kennicutt's empirical rule.

It must be stressed, however, that we have not explained why the value
$\sigma = 6 \hbox{ km s}^{-1}$ should be so universal for disk galaxies.
Linking it to the Alfv\'en velocity implies a magnetic field of about
3~$\mu$G -- close to the field value inferred for many galaxies.
Our achievement is to show that MHD instabilities can then maintain
turbulent velocity fluctuations of the order of the the Alfv\'en speed,
but why this velocity should always be $\sim 6 \hbox{ km s}^{-1}$ is
clearly related to why the magnetic field is always a few microgauss.
This, we have yet to understand.

\section*{Acknowledgments}

We are grateful to the Isaac Newton Institute for Mathematical Sciences
and its director, Keith Moffatt, for hospitality and support while this
work was largely completed.  It is a pleasure to acknowledge conversations
with R.\ Chevalier, A.\ Ferguson, C.\ Gammie, J.\ Hawley, R.\ Sancisi,
and S.\ Sridhar.  This work was supported by NSF grant AST-96/17088 and
NASA grant NAG 5-2803 to JAS and by NSF grant AST-94/23187 and NASA grants
NAG-5-3058 \& NAGW-4431 to SB.

%\ifodd\style \vfill\eject \phantom{junk} \fi
\vfill\eject

\begin{table}
\setlength{\baselineskip}{12pt}
\def\fm{\footnotemark}
\setlength{\baselineskip}{12pt}
\centering
\begin{tabular}{lcc}
   \multicolumn{3}{c}{Table 1. 
    Adopted properties in NGC 1058} \\ \hline \hline
   Radius \dotfill  &  5 kpc ($100^{\prime\prime}$)
                    & 10 kpc ($200^{\prime\prime}$) \\ \hline
   \HI\ column density \dotfill & $6 \times 10^{20} \hbox{ cm}^{-2}$ & 
                                  $4 \times 10^{20} \hbox{ cm}^{-2}$  \\
   Total gas surface density $\Sigma_{g}$ 
                              & $1.5\times 10^{-3} \hbox{ g cm}^{-2}$
                              & $10^{-3} \hbox{ g cm}^{-2}$  \\
   Full vertical thickness $h$ \dotfill & 400 pc & 400 pc \\
   Mean gas density $\bar\rho$ \dotfill &
                              $1.2\times 10^{-24} \hbox{ g cm}^{-3}$ &
                              $0.8\times 10^{-24} \hbox{ g cm}^{-3}$ \\
   Circular velocity $V_{\rm circ}$ \dotfill & 150 km s$^{-1}$
                                             & 150 km s$^{-1}$ \\
   Velocity dispersion in the gas $\sigma$ & 6 km s$^{-1}$ & 6 km s$^{-1}$ \\
   Epicyclic frequency $\kappa$ \dotfill & $1.4 \times 10^{-15}\hbox{ s}^{-1}$
                                       & $0.7 \times 10^{-15}\hbox{ s}^{-1}$ \\
   Local stability parameter $Q_g$ \dotfill  & 2.7 & 2 \\
   Jeans length $\lambda_{rm crit}$ & 670 pc & 1,800 pc \\ \hline \hline
\label{tab:one}
\end{tabular}
\end{table}


\begin{references}

\reference{}
Balbus, S.\ A.\ \& Hawley, J.\ F.\ 1991, \apj, {\bf 376}, 214

\reference{}
Balbus, S.\ A.\ \& Hawley, J.\ F.\ 1998, Rev. Mod. Phys., {\bf
70}, 1

\reference{}
Beck, R., Brandenburg, A., Moss, D., Shukurov, A.\ \& Sokoloff, D.\ 1996,
\araa, {\bf 34}, 155

\reference{}
Binney, J., \& Tremaine, S. 1987, {\it Galactic Dynamics},
(Princeton: Princeton Univ. Press)

\reference{}
Boulanger, F.\ \& Viallefond, F.\ 1992, \aap, {\bf 266}, 37

\reference{}
Braun, R.\ 1997, \apj, {\bf 484}, 637

\reference{}
Braun, R.\ 1998, astro-ph/9804320

\reference{}
Bregman, J.\ 1980, \apj, {\bf 236}, 577

\reference{}
Carlberg, R.\ G.\ \& Sellwood, J.\ A.\ 1985, \apj, {\bf 292}, 79

\reference{}
Caselli, P.\ \& Myers, P. 1995, \apj, {\bf 446}, 665

\reference{}
Dickey, J.\ M. 1996  In IAU Symp 169 p~489

\reference{}
Dickey, J.\ M., Hanson, M.\ M.\ \& Helou, G.\ 1990, \apj, {\bf 352} 522

\reference{}
Dickey, J.\ M., \& Lockman, F.\ J.\ 1990, \araa, {\bf 28}, 215

\reference{}
Eardley, D.\ M., \& Lightman, A.\ P. 1975, \apj, {\bf 200},
187

\reference{}
Ferguson, A.\ M.\ N.\ 1997, PhD Thesis, Johns Hopkins University

\reference{}
Ferguson, A.\ M.\ N., Wyse, R.\ F.\ G., Gallagher, J.\ S.\ \& Hunter, D.\ 1998,
ApJL submitted

\reference{}
Gammie, C.\ F.\ \& Ostriker, E.\ C.\ 1996, \apj, {\bf 466}, 814

\reference{}
Hawley, J.\ F.\,  Gammie, C.~F., \& Balbus, S.\ A. 1995, \apj,
{\bf 440}, 742

\reference{}
Heiles, C.\ 1996, in ``Polarimetry of the Interstellar Medium,'' eds.\
W.\ G.\ Roberge \& D.\ C.\ B.\ Whittet (San Franciso, ASP) vol.\ 97, p.~457

\reference{}
Heiles, C., Goodman, A., McKee, C.\ \& Zweibel, E.\ 1993, in
``Protostars and Planets III,'' eds.\ E.\ H.\ Levy \& J.\ I.\ Lunine
(Tuscon: U. Arizona Press) p.~279

\reference{}
Jenkins, A.\ \& Binney, J.\ 1990, \mnras, {\bf 245}, 305

\reference{}
Kamphuis, J.\ 1993, PhD thesis, University of Groningen

\reference{}
Kamphuis, J.\ \& Sancisi, R.\ 1993, \aap, {\bf 273}, L31

\reference{}
Kennicutt, R.\ C.\ 1989, \apj, {\bf 344}, 685

\reference{}
Kulkarni, S.\ R.\ \& Fich, M.\ 1985, \apj, {\bf 289}, 792

\reference{}
Leitherer, C.\ \& Heckman, T.\ M.\ 1995, \apjs, {\bf 96}, 9

\reference{}
Lynden-Bell, D., \& Kalnajs, A.\ J.\ 1972, \mnras, {\bf 157},
1

\reference{}
McKee, C.\ F.\ \& Zweibel, E.\ G.\ 1995, \apj, {\bf 440}, 686

\reference{}
Miesch, M.\ S.\ \& Bally, J.\ 1994, \apj, {\bf 429}, 645

\reference{}
Norman, C.\ A.\ \& Ferrara, A.\ 1996, \apj, {\bf 467}, 280

\reference{}
Norman, C.\ A.\ \& Ikeuchi, S.\ 1989, \apj, {\bf 345}, 372

\reference{}
Radhakrishnan, V.\ \& Srinivasan, G.\ 1980, J.\ Astr.\ Ap., {\bf 1}, 47

\reference{}
Rownd, B.\ K., Dickey, J.\ M.\ \& Helou, G.\ 1994, \apj, {\bf 108}, 1683

\reference{}
Scalo, J.\ M.\ 1987, in {\it Interstellar Processes}, eds. D.\
J.\ Hollenbach \& H.\ A. Thronson (Dordrecht: Reidel), 349

\reference{}
Schulman, E., Bregman, J.\ N., Brinks, E.\ \& Roberts,
M.\ S.\ 1996, \aj, {\bf 112}, 960

\reference{}
Spitzer, L.\ 1978, {\it Physical Processes in the ISM}, (New
York: Wiley)

\reference{}
Tamman, G.\ A.\ 1974, in {\it Supernovae and Supernova Remnants}, ed.\
C.\ B.\ Cosmovici (Dordrecht: Reidel) p.~215

\reference{}
Toomre, A.\ \& Kalnajs, A.\ J.\ 1991, in {\it Dynamics of Disc Galaxies},
ed.\ B.\ Sundelius (Gothenburg: G\"oteborgs University) p.~341

\reference{}
T\'oth, G.\ \& Ostriker, J.\ P.\ 1992, \apj, {\bf 389}, 5

\reference{}
van der Kruit, P.\ C.\ \& Shostak, C.\ S.\ 1984, \aap, {\bf 134}, 258

\reference{}
Wolfire, M.\ G., Hollenbach, D., McKee, C.\ F., Tielens, A.\ G.\ G.\ \& Bakes,
E.\ L.\ O.\ 1995, \apj, {\bf 443}, 152

\end{references}
\end{document}